\documentclass{format170x240multiauthor}
\usepackage[T1]{fontenc}
\usepackage{makeidx}
   \makeindex
\usepackage{amsmath}
\usepackage{textcomp}
\usepackage{graphicx}
\usepackage{cite} 

\usepackage{gensymb}
\usepackage{units}
\usepackage{isotope}

\newcommand{\ket}[1]{\left|  #1 \right\rangle}

\newcommand{\aver}[1]{\left\langle {#1} \right\rangle}
\newcommand{\Alion}{\ensuremath{\mathrm{Al}^+}}
\newcommand{\Cs}{\isotope[133]{Cs}}
\newcommand{\Rb}{\isotope[87]{Rb}}
\newcommand{\Sr}{\isotope[87]{Sr}}
\newcommand{\avqzshift}{\ensuremath{\aver{\delta \nu}_Z}}
\newcommand{\collshift}{\ensuremath{\delta \nu_C}}
\newcommand{\avcollshift}{\ensuremath{\aver{\delta \nu}_C}}
\newcommand{\omegabar}{\ensuremath{\bar{\omega}}}

\begin{document}

\chapter{Microchip-based trapped-atom clocks}
\label{Vuletic:chap}
\chapterauthor{Vladan Vuleti\'{c}} 
\chapterauthor{Ian D. Leroux} 
\chapterauthor{Monika H. Schleier-Smith} 

\section{Basic principles}
\label{Vuletic:sec:ClockBasics}

A two-level quantum system in vacuum---in the absence of any perturbing fields---constitutes an ideal clock whose oscillation frequency $\omega = E/\hbar$ is given by the energy difference $E$ between the two levels $\ket{1},\ket{2}$. \index{atomic clock}
In a single measurement of duration $T$ performed on a single particle this frequency can be determined with an uncertainty $\Delta \omega = 1/T$.
If the measurement is performed simultaneously on $N$ independent identical particles, and this measurement is repeated with a cycle time $T_c>T$ for a total averaging time $\tau > T_c$, the fundamental quantum uncertainty of the frequency determination \index{standard quantum limit} is given by the standard quantum limit \cite{Vuletic:Santarelli99}
\begin{equation}\label{Vuletic:eq:Precision}
    \frac{\Delta \omega}{\omega} = \frac{1}{\omega T}\sqrt{\frac{T_c}{N\tau}}.
\end{equation}
The $N^{-1/2}$ scaling arises because the quantities to be measured are the non-zero probabilities to find a particle in either of the clock states $\ket{1}, \ket{2}$: when the independent particles in the ensemble are read out, the observed populations of the two clock states are binomially distributed, leading to so-called projection noise on the estimation of those probabilities \cite{Vuletic:Wineland92,Vuletic:Wineland94}.  The $\sqrt{T_c/\tau}$ scaling arises because the sequential measurement repeated $\tau/T_c$ times using $N$ particles each is equivalent to a single measurement using $N \tau/T_c$ atoms.  For a given total measurement time $\tau$ the stability improves as the duration $T$ of the single measurement is increased.  The latter is limited by the coherence time of the transition.  While the absolute stability does not depend on the transition frequency $\omega$, the fractional stability improves with higher transition frequency.

\section{Atomic-fountain versus trapped-atom clocks}
\label{Vuletic:sec:TrappedAtomClocks}

In the absence of other fields, particles fall under gravity, which sets a practical limit on the single-measurement time $T$.  Atomic fountains \index{fountain clock} \cite{Vuletic:Kasevich89,Vuletic:Wynands05}, where an ensemble of atoms is launched upwards into a ballistic flight region, allow one to increase the measurement time. \index{atomic fountain}  A vacuum apparatus with a height of \unit[1]{m} yields a typical interaction time of $T\sim\unit[0.7]{s}$.
While larger systems are under construction for the measurement of weak gravitational effects\cite{Vuletic:Dimopoulos07}, substantial further increase of measurement time for freely falling atoms is impractical---unless working in a microgravity environment \cite{Vuletic:Laurent06,Vuletic:Esnault07}---in view of the quadratic dependence of apparatus height on measurement time.

\begin{figure}[htb]
\includegraphics[width=3.5in]{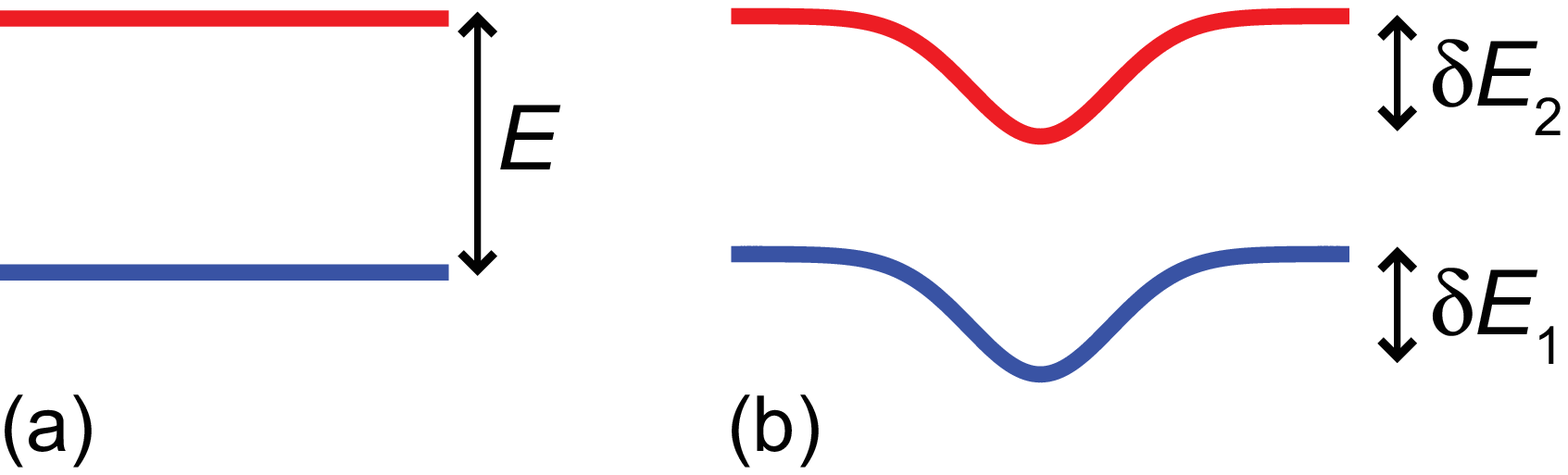}
\Caption{Illustration of free-space clock (A) and trapped-atom clock (B). A trapped atom clock can be operated with small linewidth if the difference $\delta E = \delta E_2 - \delta E_1$ between the energy shifts $\delta E_1,\delta E_2$ of the trapped states is small.}
\label{Vuletic:fig:CommonShift}
\end{figure}

Achieving long measurement times $T$ with reasonable-sized apparatus therefore requires the use of trapped atoms, held against gravity by some externally applied force.
Such a force necessarily perturbs the atomic energy levels $E_1$ and $E_2$, but this perturbation can be tolerated provided that the \emph{differential} energy shift between the two clock levels $\delta E=\delta E_1- \delta E_2$ is sufficiently small (see Fig. \ref{Vuletic:fig:CommonShift}).
Under such circumstances the external field can be used to provide a trap for the particles without compromising clock stability or accuracy.
\index{magic trap}
An example is the electrostatic Coulomb force used to trap ions, which produces only a small differential shift of the clock transition \cite{Vuletic:Bollinger91,Vuletic:Rosenband08}.
In such systems, a coherence time of \unit[10]{min} has been reported between hyperfine ground states using $\mathrm{Be}^+$ ions \cite{Vuletic:Bollinger91}.
To date, a single \Alion\ ion trapped in a Paul trap constitutes the best clock, with current fractional accuracy now exceeding $10^{-17}$ using an optical transition \cite{Vuletic:Chou10,Vuletic:Rosenband08}.
Similar non-perturbing traps can be crafted for hyperfine transitions in neutral atoms using magnetostatic forces \cite{Vuletic:Migdall85,Vuletic:Bagnato87,Vuletic:Hess87} provided the linear Zeeman shift is the same in both clock states \cite{Vuletic:Harber02,Vuletic:Treutlein04}.
For electronic transitions at optical frequencies traps based on the optical dipole force (AC Stark Shift) \cite{Vuletic:Chu86,Vuletic:Grimm00} can be used at certain ``magic wavelengths'' where the polarizability of the two clock states is the same \cite{Vuletic:Taieb94}.
This approach, developed independently for studies of cavity quantum electrodynamics by Kimble and coworkers \cite{Vuletic:Kimble99,Vuletic:McKeever03} and for optical-transition clocks by Katori and coworkers \cite{Vuletic:Katori99,Vuletic:Ido00,Vuletic:Katori03}, has enabled many of the recent successes using magic-wavelength optical traps for neutral atoms \cite{Vuletic:Takamoto03,Vuletic:Takamoto05,Vuletic:Santra05,Vuletic:Ido05,Vuletic:Boyd06,Vuletic:Boyd07,Vuletic:Blatt08,Vuletic:Ye08,Vuletic:Blatt08,Vuletic:Campbell09,Vuletic:Lodewyck09}

Aside from allowing long interrogation times, the confinement of atoms to a small volume, typically of millimeter to micrometer size, allows better control of perturbing external fields than can be achieved over the meter-scale flight region of an atomic fountain. \index{collision shift} \index{density shift}
On the other hand, this confinement also leads to a higher atomic density for a given atom number, resulting in larger collision shift of the clock transition frequency than in the dilute clouds used in fountain clocks.
These collision shifts can be suppressed for trapped-atom clocks at the expense of an increase in overall trapping volume, by choosing periodic confining potentials with less than one atom per site so that atoms never collide \cite{Vuletic:Takamoto05,Vuletic:Santra05,Vuletic:Ido05,Vuletic:Boyd06,Vuletic:Boyd07,Vuletic:Blatt08,Vuletic:Ludlow08}.
However, even without such a suppression the achievable accuracy of trapped-atom clocks is interesting for many commercial applications.
Microchip-based atom traps, which allow compact experimental setups with modest power requirements, might thus allow the construction of robust, portable trapped-atom secondary frequency standards that would be technologically valuable even if they do not exceed the absolute performance of fountain clocks in the laboratory~\cite{Vuletic:Treutlein04,Vuletic:Rosenbusch09}.
The hope of making the stability of cold-atom clocks available in the field has fueled much of the commercial and experimental interest in chip clocks.

\section{Optical-transition clocks vs microwave clocks}
\label{Vuletic:sec:OpticalClocks}

Since the fractional accuracy of an atomic clock improves with increasing transition frequency, it is natural to consider the use of optical-frequency electronic transitions rather than microwave-frequency hyperfine transitions as the basis for a clock. \index{optical clock} \index{magic wavelength}
Magic-wavelength clocks operating on optical transitions have already surpassed the much more mature hyperfine-transition atomic fountain clocks in accuracy \cite{Vuletic:Takamoto03,Vuletic:Takamoto05,Vuletic:Santra05,Vuletic:Ido05,Vuletic:Boyd06,Vuletic:Boyd07,Vuletic:Blatt08,Vuletic:Ludlow08,Vuletic:Ye08}, and further improvements are expected.

An optical clock could be constructed on an atom chip, using integrated fiber optics for the optical trapping and probing fields \cite{Vuletic:Colombe07}.
However, the measurement of optical transition frequencies currently requires bulky and vibration-sensitive laser systems with ultra-stable reference cavities and optical frequency combs, negating the advantages of compactness and robustness that make chip clocks interesting in the first place. It is therefore likely that work on chip clocks will concentrate, for the medium term, on hyperfine transitions in the microwave region of the electromagnetic spectrum.

\section{Clocks with magnetically trapped atoms: fundamental limits to performance}
\label{Vuletic:sec:MagneticTrapClocks}

Microwave clocks cannot use the same optical dipole traps as optical-transition clocks due to the lack of suitable ``magic wavelengths'' for hyperfine transitions (but see reference \cite{Vuletic:Flambaum08} for a recently proposed workaround which exploits the tensor polarizability).
Instead, a ``magic''-confinement approach is available for hyperfine clocks using magnetic trapping.
While fountain clocks use magnetically untrapped $\ket{F,m_F=0},\ket{F+1,m_F=0}$ states for which the linear Zeeman shift vanishes at zero magnetic field, it is also possible to find magnetically trappable states where the \emph{difference} between the Zeeman shift for the two states $\ket{1},\ket{2}$ for some magnetic-field value $B_0$ varies only quadratically around $B_0$.
Thus it is possible to realize a hyperfine clock with magnetically trapped atoms that has similar sensitivity to external magnetic fields as a fountain clock \cite{Vuletic:Harber02,Vuletic:Treutlein04,Vuletic:Rosenbusch09}. \index{magnetic-trap clock transition}

\begin{figure}[htb]
\includegraphics[width=3.5in]{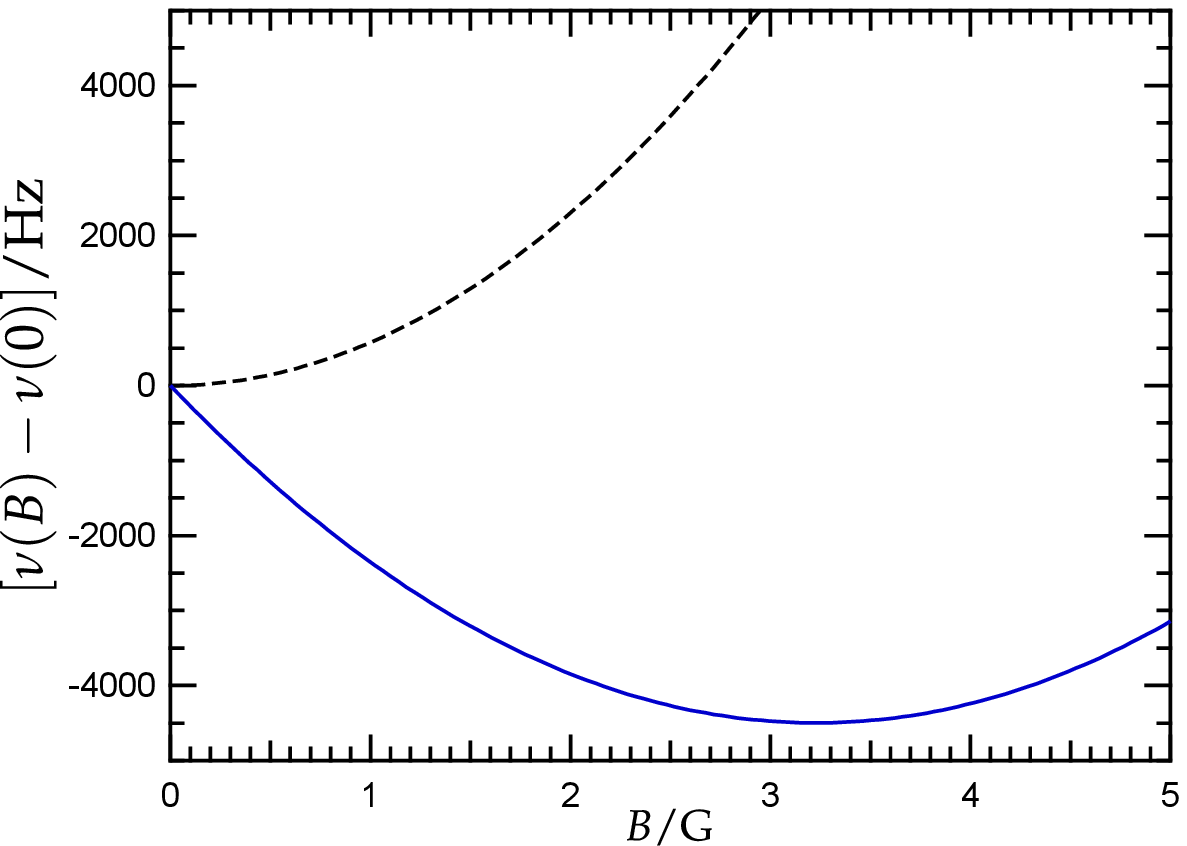}
\Caption{Transition frequency shift vs magnetic field for the standard clock transition
$\ket{F=1,m_F=0} \rightarrow \ket{F=2,m_F=0}$ (dashed line), and for the  ``magic''
magnetic-trap transition $\ket{F=1,m_F=-1} \rightarrow \ket{F=2,m_F=1}$ in $^{87}$Rb. For the former,
there is no linear Zeeman shift at zero magnetic field, while for the latter, the linear Zeeman shift
vanishes at the field of $3.23$~G.}
\label{Vuletic:fig:FrequencyVsB0}
\end{figure}

Fig. \ref{Vuletic:fig:FrequencyVsB0} shows the dependence of the hyperfine transition frequency on magnetic fields for a standard $^{87}$Rb fountain clock that uses the transition $\ket{F=1,m_F=0} \rightarrow \ket{F=2,m_F=0}$, and for a trapped-atom clock that uses the transition $\ket{1} \equiv \ket{1,-1} \rightarrow \ket{2} \equiv \ket{2,1}$. For the latter, the coefficient for the quadratic variation
\begin{equation}\label{Vuletic:eq:FreqShift}
    \nu(B)=\nu_0+\beta (B-B_0)^2
\end{equation}
around the magic field $B_0=\unit[3.228917(3)]{G}$ is given by $\beta = \unit[431.35957(9)]{Hz/G^2}$ for \Rb\ \cite{Vuletic:Harber02}, which is a little smaller than for the fountain-clock transition $\ket{1,0} \leftrightarrow \ket{2,0}$, where it amounts to $\beta' = \unit[575.14]{Hz/G^2}$ \cite{Vuletic:QPAFS}. The transition frequency at the minimum of the $\ket{1,-1} \leftrightarrow \ket{2,1}$ transition is given by $\nu_0=\unit[6 834 678 113.59(2)]{Hz}$ for \Rb\ \cite{Vuletic:Lewandowski02}, slightly smaller than the hyperfine splitting and transition frequency of the $\ket{1,0} \leftrightarrow \ket{2,0}$ transition at zero field, $\nu_0'= \unit[6 834 682 610.904 32(2)]{Hz}$ \cite{Vuletic:Wynands05}.

However, in a fountain clock one can apply a uniform magnetic field, whereas the magic-magnetic-trap approach requires the atoms to experience spatially varying magnetic-field magnitude for trapping. \index{collision shift} \index{density shift}
This means that the thermal motion of the atoms in the trap will necessarily cause them to sample regions of the potential with different magnetic fields, and thus different residual quadratic Zeeman shift.
The range of magnetic fields sampled by the atoms, and the resulting broadening and shift of the clock transition, increase as the cloud's temperature increases, so it seems advantageous to operate the clock with an ensemble of atoms that is as cold as possible.
However, cooling the sample increases its density (for fixed atom number) and thus leads to a higher collision shift.
Since the two perturbations have opposite temperature dependences, a compromise must be found.
Here we analyze these two dominant line shift mechanisms, following the arguments of references~\cite{Vuletic:Harber02,Vuletic:Rosenbusch09}, in order to find the achievable performance for a magnetic-trap hyperfine clock as first demonstrated by Harber and coworkers in a macroscopic trap \cite{Vuletic:Harber02}, pioneered for a microchip trap by Treutlein \emph{et al.} \cite{Vuletic:Treutlein04} and recently upgraded to a precision device by Ram\'{i}rez-Mart\'{i}nez \emph{et al.} \cite{Vuletic:Ramirez-Martinez10}.

We restrict the analysis to a magnetically-trapped \Rb\ hyperfine clock and do not consider a clock using \Cs. First, unlike \Rb\ where the density shift is independent of temperature for the temperature range of interest (i.e., temperatures lower than $\unit[\sim100]{\micro K}$), the density shift for \Cs\ remains temperature-dependent down to \unit{nK} temperatures \cite{Vuletic:Leo01} due to a multitude of low-field Feshbach resonances \cite{Vuletic:Vuletic99,Vuletic:Chin00,Vuletic:Leo00,Vuletic:Kerman01,Vuletic:Chin03,Vuletic:Chin04}.
Related, but more important, is the fact that the density shift in \Cs, at typical temperatures of interest, is about two orders of magnitude larger than in \Rb, so that microchip magnetic traps, with their relatively high atomic densities, would yield very poorly performing \Cs\ clocks with large line shift and broadening.

As an atom moves in the trap, it experiences a time-varying magnetic field $B(x(t))$ that constitutes the source of its potential energy $U(B(x))$. This potential energy in a trap with a field minimum $B_\text{min}$ is given by
\begin{equation}\label{Vuletic:eq:Potential}
   U(B)= g_F m_F \mu_B (B-B_\text{min}) = \Upsilon (B-B_{min}),
\end{equation}
where $g_F$ is the atomic Land\'{e} factor, $\mu_B$ the Bohr magneton, and $\Upsilon/h \approx
\unit[0.70]{MHz/G}$ for \Rb\ for the trapped states $\ket{F=1,m_F=-1}, \ket{F=2,m_F=1}$ \cite{Vuletic:QPAFS}.

It then follows that if we choose the minimum trap field equal to the magic field, $B_\text{min}=B_0$, the average quadratic Zeeman shift of the clock transition is given by \index{quadratic Zeeman shift}
\begin{equation}\label{Vuletic:eq:AverageQuadZeemanShift}
   \avqzshift = \frac{\beta}{\Upsilon^2} \aver{U^2}.
\end{equation}
For a thermal distribution of temperature $T$ in a three-dimensional harmonic trap the clock frequency shift can be written as
\begin{equation}\label{Vuletic:eq:AverThermalZeemanShift}
    \avqzshift|_T = \frac{15 \beta}{4 \Upsilon^2} (k_B T)^2 = \zeta T^2.
\end{equation}
Thus we find that the clock Zeeman shift is quadratic in the atomic temperature with a coefficient $\zeta =
\unit[1.43]{Hz/\micro K^2}$ for \Rb.
We have assumed that the temperature is large compared to the trap oscillation frequency, so that the trap zero-point energy can be ignored. This is a good approximation for the optimum range of temperatures once the density shift is taken into account.

The Zeeman frequency shift and broadening due to the thermal energy of the gas in the trap must be traded off with the collision shift, which for fixed atom number and trap frequency increases as  the atoms get colder. For a given atomic density $n$ and equal populations of the two clock states the frequency shift is given by the expression \cite{Vuletic:Harber02,Vuletic:Gibble09,Vuletic:Rosenbusch09} \index{collision shift} \index{density shift}
\begin{equation}\label{Vuletic:eq:CollFreqShift}
    \collshift = \frac{2 \hbar}{m} n \left( a_{22}-a_{11}  \right).
\end{equation}
Here $a_{22}$ and $a_{11}$ are the scattering lengths for $\ket{2}+\ket{2}$ and $\ket{1}+\ket{1}$ collisions, respectively. Thus the frequency shift is proportional to the atomic density and the difference in scattering lengths for the different clock states. For \Rb\ a serendipitous degeneracy leads to almost identical scattering lengths $a_{11}=100.44 a_0$ and $a_{22}=95.47 a_0$ \cite{Vuletic:Lewandowski02,Vuletic:Vogels00}, where $a_0$ is the Bohr radius.

The density shift for a thermal cloud of $N$ atoms in a harmonic trap can be written as
\begin{equation}\label{Vuletic:eq:RbCollFreqShift}
    \avcollshift = \frac{2 \hbar}{m} \left( a_{22}-a_{11}  \right) \aver{n} = - \chi \frac{N \omegabar^3} {T^{3/2}}
\end{equation}
with $\omegabar=\sqrt[3]{\omega_x \omega_y \omega_z}$ the geometric mean of the trap frequencies along the three axes, and $\chi = \unit[9.2\times 10^{-15}]{Hz~s^3~\micro K^{3/2}}$.
\begin{figure}[htb]
\includegraphics[width=3.5in]{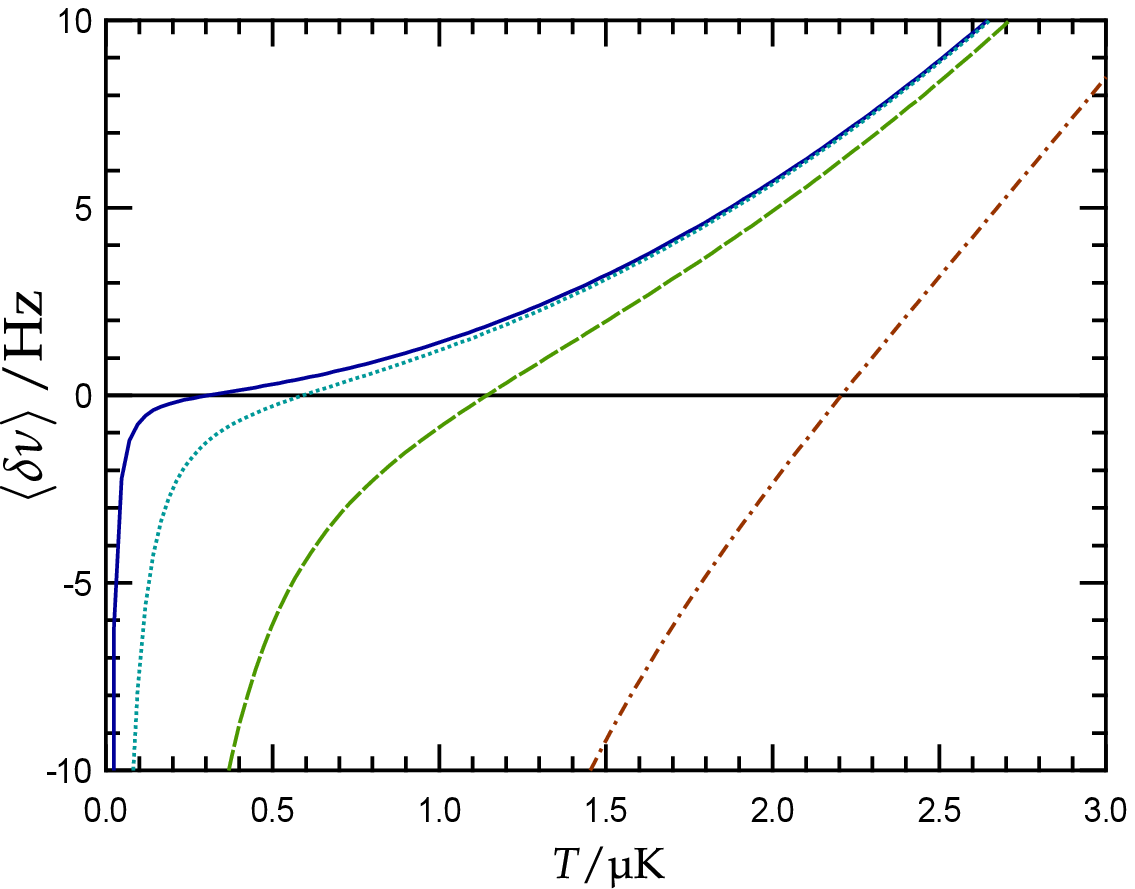}
\Caption{The total (Zeeman and collision) frequency shift of the magic-field clock transition $\ket{1,-1} \rightarrow \ket{2,1}$ for an ensemble of magnetically trapped $^{87}$Rb atoms  vs. ensemble temperature. A trap with a vibration frequency $\omegabar = 2 \pi \times 100$~Hz is assumed. The curves from left to right correspond to $10^4, 10^5, 10^6$, and $10^7$ trapped atoms.}
\label{Vuletic:fig:TotalShift}
\end{figure}

To maximize coherence time in the magnetic trap, it is convenient to choose parameters such that the Zeeman shift and collision shift are of the same magnitude.
Since they have approximately the same position dependence in the trap, the inhomogeneous shifts which they respectively impose on the cloud will then cancel each other, considerably extending the coherence time of the Ramsey fringes \cite{Vuletic:Rosenbusch09}.
We therefore consider clock operation at temperature such that \avqzshift\ and \avcollshift\ are of the same magnitude and therefore sum to zero.
Fig. \ref{Vuletic:fig:TotalShift} shows the combined quadratic Zeeman shift and density shift of the clock transition as a function of temperature for different total trapped atom numbers. For $N=10^6$ atoms in a trap with vibration frequencies $\omega_x =\omega_y = \omega_z = \unit[2 \pi \times 100]{Hz}$ the shift goes through zero at a temperature of $T_0=(\chi\omegabar^3N/\zeta)^{2/7} =  \unit[1.1]{\micro K}$.
Under these conditions the rms cloud size in each dimension is $\sqrt{k_B T_0/m}\omegabar^{-1} = \unit[17]{\micro m}$ and the shift from each effect individually is $\unit[1.9]{Hz}$.
Note that $T_0/T_c = (\unit[0.3]{mK}/T_c)^{1/7}$, where $k_B T_c = 0.94\hbar\omegabar N^{1/3}$ is the critical temperature for Bose Einstein condensation\cite{Vuletic:BECDG}.  For the (low) densities of interest for clock operation, $T_c \ll \unit[0.3]{mK}$ so that the zero-shift temperature $T$ is always above the critical temperature, and the clock always stays in the classical (non-degenerate) regime.
Assuming that the atom number can be measured and controlled to 5 \%, a fractional accuracy of $10^{-11}$ on the hyperfine transition at \unit[6.8]{GHz} appears possible in such a trap.
A more detailed analysis, including the effects of motional and collisional averaging \cite{Vuletic:Harber02,Vuletic:Deutsch10}, and assuming better control of the atom number by measuring it at the end of each experimental cycle, suggests that fractional stability of $10^{-12}$ and fractional accuracy at the level of $10^{-13}$ might be attainable \cite{Vuletic:Rosenbusch09}.

Such accuracy is much worse than the state of the art in hyperfine atomic fountains that have reached $10^{-16}$ fractional accuracy~\cite{Vuletic:Wynands05}.
Nevertheless, the small size of the microchip clock in combination with its accuracy renders it interesting as a secondary standard for commercial or other applications.
Also it may be possible to improve upon the parameters discussed above.
For instance, a trap with a vibration frequency as low as \unit[10]{Hz} has been demonstrated in the study of quantum reflection from a surface \cite{Vuletic:Pasquini06}.  In such a trap and with $N=10^5$ atoms, the optimum temperature is around $\unit[80]{nK}$, reducing the quadratic Zeeman and collision shifts to below \unit[10]{mHz}.

The density shift could be avoided altogether if the atoms were confined in an array of traps, such that at most one atom is trapped at each individual trap, as with optical lattices.  It is possible to make an array of magnetic traps on the microchip \cite{Vuletic:Gerritsma07}, but the individual trap size is larger than for optical lattices, resulting in a larger total trap volume. Nevertheless, with a 30~$\mu$m lattice period a planar array of $10^5$ traps could be created within a $\unit[1]{cm^2}$ area. \index{array of traps}

An additional challenge that is specific to microchip clocks is that they are operated in close vicinity to the chip surface, where electric and magnetic fields are enhanced.
Electric-field effects are negligible for the hyperfine spin states of interest for clock operation (see Chapter ...), while magnetic-field effects are understood \cite{Vuletic:Henkel99,Vuletic:Henkel01} and have been quantified in experiments (see Chapter ...).
In particular, there are increased magnetic field fluctuations near conducting surfaces that are due to Johnson-noise-induced currents in the conductor \cite{Vuletic:Lin03}.
However, such fields can be reduced by using a chip design where the trap is located close to a non-conducting surface, but at relatively large distance from any conductor \cite{Vuletic:Lin03}.

\section{Clocks with magnetically-trapped atoms: experimental
  demonstrations}
\label{Vuletic:sec:MagneticTrapClockExpts}

The tradeoff between Zeeman shift and collision shift was investigated experimentally by Eric Cornell's group at Boulder in a standard setup for the creation of Bose-Einstein condensates, a macroscopic magnetic trap \cite{Vuletic:Harber02}. An ensemble of typically $10^6$ \Rb\ atoms was prepared by a combination of laser and evaporative cooling at a typical temperature of \unit[500]{nK} for thermal clouds, or as a Bose-Einstein condensate with large condensate fraction. Starting from the state $\ket{1}=\ket{F=1,m_F=-1}$, the atoms were then prepared in a superposition of states $\ket{1}$ and $\ket{2} \equiv \ket{F=2,m_F=1}$ using a two-photon microwave transition. The transition frequency and clock coherence were then investigated using the Ramsey technique of separated oscillatory fields as a function of both atomic density and offset magnetic field $B_0$ at the bottom of the magnetic trap.

For the transition frequency $\nu_{12}$ as a function of magnetic field, the expected quadratic variation of equation \ref{Vuletic:eq:FreqShift} was observed with a quadratic coefficient $\beta \approx \unit[431]{Hz/G^2}$.
In such a setup, control of the magnetic field to a level of $10^{-3}$ (\unit[3]{mG}) is straightforward, which in itself would permit a clock accuracy of $\unit[4]{mHz}$, corresponding to a fractional stability of $10^{-12}$.
With magnetic shielding in combination with interspersed measurement of the magnetic field on a transition with linear Zeeman shift, and corresponding field compensation, one would expect to be able to achieve a field stability on the order of \unit[10]{\micro G}, which would limit the clock accuracy only at the level of $10^{-17}$.
However, as as discussed in section~\ref{Vuletic:sec:MagneticTrapClocks}, collision shifts limit the clock accuracy long before this level is reached.

Section~\ref{Vuletic:sec:MagneticTrapClocks} shows that the quadratic coefficient $\beta$ also quantifies the frequency shift and decoherence that arise from the sample's small, but non-zero temperature. For the trap parameters of the experiment by Harber \emph{et al.} the quadratic Zeeman shift of the clock transition amounted to $\unit[\sim 1]{Hz}$ at a sample temperature of $\unit[500]{nK}$.
For low-density clouds coherence times in excess of $\unit[2]{s}$ were observed, which demonstrated that in terms of coherence times trapped atoms can compete with atomic fountains.
More recently, coherence times approaching one minute have been observed with atoms trapped on a microchip by a collaboration led by Jakob Reichel (\'{E}NS) and Peter Rosenbusch (SYRTE)~\cite{Vuletic:Deutsch10}.

The relatively high density in a trapped-atom clock then makes collision shifts easily observable. The density shift becomes particularly large when a Bose-Einstein condensate is formed, at much larger density than the associated thermal cloud. For the experiment by Harber \emph{et al.} with trap vibration frequencies $(\omega_x, \omega_y, \omega_z)$ of $2 \pi \times (230,230,7)$~Hz, the transition frequency shift in a non-degenerate gas at a temperature of $T=500$~nK amounted to $-3.9(3) \times 10^{-13}$~Hz cm$^3$, or typically $\aver{\delta \nu}_C =-5$~Hz. For the same trap parameters the density shift in the pure condensate with $N=10^6$ atoms was as large as $\aver{\delta \nu}_C = -25$~Hz. \index{density shift in a BEC}

An intriguing feature of the collision shift in a degenerate quantum gas is the factor of two appearing in equation \ref{Vuletic:eq:CollFreqShift}, associated with exchange symmetry for bosons \cite{Vuletic:Harber02}, and a similar effect for fermions that could be dubbed the 'factor of (not) zero' \cite{Vuletic:Campbell09}. (See also the work by Kurt Gibble \cite{Vuletic:Gibble09} for a unified approach for bosons \cite{Vuletic:Killian98,Vuletic:Harber02} and fermions \cite{Vuletic:Oktel02,Vuletic:Zwierlein03}.) For identical bosons, the $s$-wave collision cross section is given in terms of the $s$-wave scattering length $a$ by by $\sigma_\text{id} = 8 \pi a^2$, whereas the corresponding expression for distinguishable particles (e.g., bosons in different quantum states) with the same scattering length is two times smaller, $\sigma_\text{dist} = 4 \pi a^2$. The difference arises from the two different possible paths in the collision that give rise to the same final state, su!
 ch that the amplitudes for the processes add, giving the larger value for identical bosons.
For a coherently prepared sample, all particles are in the same internal state, and the larger cross section for identical bosons applies, as verified experimentally by Harber \emph{et al.} \cite{Vuletic:Harber02}. \index{exchange symmetry}

Another curious observation made in this experiment was that collisions actually serve to lengthen the coherence time of the sample.
Indeed, the observed decay time of the Ramsey oscillations could be up to eight times longer than what would have been predicted based on the quadratic Zeeman broadening of the transition in a simple collision-free model.
The authors concluded that the collisions, by randomly exchanging the velocities of pairs of particles in the cloud, cause them all to sample the same magnetic field environment.
Thus all the atoms in the cloud experience substantially the same average Zeeman shift, even if the shifts are different for each atom at any given instant. \index{coherence time}
The long coherence times observed by Deutsch \emph{et al.} in their chip trap are also believed to be due in part to collisions, but the hypothesized mechanism is different: under certain conditions colliding identical atoms can coherently exchange their phase, so that an atom that has precessed farther than its collision partner finds its phase set back, rather as though a spin-echo procedure were continually being applied to the sample \cite{Vuletic:Rosenbusch09,Vuletic:Deutsch10}.

The experiment by Harber \emph{et al.} \cite{Vuletic:Harber02} demonstrated a Ramsey clock with cold atoms and verified the parameters determining clock performance, but used a macroscopic magnetic trap rather than a microchip.
Two years later, Treutlein \emph{et al.} applied this approach to microchips, demonstrating a clock with a coherence time of \unit[1]{s}, and verified that clock operation was possible within \unit[5]{\micro m} of the chip surface \cite{Vuletic:Treutlein04}.

Treutlein \emph{et al.} used a silver-coated microchip where the magnetic trap was loaded from a mirror magneto-optical trap (MOT), as in the first demonstration of Bose-Einstein condensation on a microchip \cite{Vuletic:Hansel01,Vuletic:Ott01}.
The loading was followed by RF-induced evaporation. In this way an ultracold atom cloud containing typically $10^4$ \Rb\ atoms was prepared at a typical temperature of $\unit[0.6]{\micro K}$ in a magic-field magnetic trap with frequencies of $\omega = \unit[2\pi\times(50, 350, 410)]{Hz}$, comparable to the example studied in section~\ref{Vuletic:sec:MagneticTrapClocks}.
The trap's position relative to the chip surface could be calculated from the known wire geometry and currents, and verified experimentally by comparing the absorption image of the atoms in the trap to the reflected image seen in the coated chip surface \cite{Vuletic:Schneider03} or, for very short distances, by studying the rate of surface-induced trap loss \cite{Vuletic:Lin03}.
The two-photon microwave transition from $\ket{F=1,m_F=-1}$ to $\ket{F=2,m_F=1}$ was driven by a microwave photon from an external antenna, detuned by $\unit[1.2]{MHz}$ from the intermediate state, and an RF photon supplied directly by an on-chip wire, yielding a Rabi frequency for the clock transition of $\unit[\sim500]{Hz}$.
The population of the clock states could then be read out by absorption imaging after a $\unit[4]{ms}$ time of flight, unfortunately with a signal-to-noise ratio of only 6, substantially worse than the projection noise limit of $\sim 100$ for the system (see section \ref{Vuletic:sec:Readout}).
To study the effect of the chip surface on atomic coherence, the authors performed Ramsey spectroscopy (i.e. operated a clock) using atomic samples trapped at varying distances from the surface.
At distances below $\unit[5]{\micro m}$, the trap loss induced by the chip surface limited the lifetime of the atomic population to a second or less, which is undesirable for clock operation.
Beyond this range, however, no effect of the chip on clock performance was observed: the coherence time---as determined from the decay of the Ramsey contrast---and the phase noise---as measured by the signal-to-noise ratio---were invariant within experimental error for atom-chip distances from $5$ to $\unit[132]{\micro m}$.

Working at a distance of $\unit[54]{\micro m}$ from the chip surface, with a $\unit[23]{s}$ experiment cycle time and a $\unit[1]{s}$ Ramsey interrogation time, Treutlein \emph{et al.} successfully operated a clock with a stability of $\sigma(\tau) = 1.7\times10^{-11} \sqrt{\unit{s}}\tau^{-1/2}$, limited by fluctuations in the quadratic Zeeman shift due to $\unit[24]{mHz}$ of magnetic field noise in the laboratory.
The long term fractional stability was limited to $\sim10^{-12}$ (after 10 minutes of integration time) by slow drifts of their microwave oscillator reference.
Since the readout noise, duty cycle and magnetic shielding can all be improved well beyond the levels used in this experiment, substantial improvements in chip clock performance beyond that demonstrated in this work are realizable.
Already, Ram\'{i}rez-Mart\'{i}nez \emph{et al.} have improved the short-term stability to $1.5\times10^{-12}\sqrt{\unit{s}}\tau^{-1/2}$ \cite{Vuletic:Ramirez-Martinez10}.

\section{Readout in trapped-atom clocks}
\label{Vuletic:sec:Readout}

The signal-to-noise ratio of clocks operating with ensembles of independent two-level atoms is limited by the projection noise associated with the independent measurement outcomes for the individual two-level atoms.
Atomic-fountain clocks operate at this limit \cite{Vuletic:Santarelli99}, while magic-wavelength optical clocks are approaching it \cite{Vuletic:Ye08}.
To achieve projection-noise-limited readout, the number of photons detected per atom must exceed one.
For absorption imaging, as used by Harber \emph{et al.} and Treutlein \emph{et al.}, it is difficult to achieve projection-noise-limited detection in view of beam intensity fluctuations and interference fringes on the camera, and both experiments were substantially above the atom projection noise limit.
For instance, in the work of Treutlein \emph{et al.}, the observed signal to noise ratio was around 6, while projection-noise-limited measurement with their $N>10^4$ atoms would have allowed a ratio of $\sqrt{N}>100$ \cite{Vuletic:Treutlein04}.
However, recent experiments have demonstrated absorption imaging at the projection noise limit \cite{Vuletic:Gross10,Vuletic:Riedel10}.

Better state readout can be obtained if the cloud is placed inside an optical resonator that serves to enhance the signal by inducing repeated atom-light interaction \cite{Vuletic:Teper06,Vuletic:Colombe07,Vuletic:Brennecke07}, or using on-chip large-aperture fiber optics \cite{Vuletic:Heine09}.
It is possible to use either fluorescence measurements or dispersive measurements, with dispersive measurements providing better signal-to-noise ratio at large atom number \cite{Vuletic:Teper06}. \index{projection noise limit} \index{optical resonator} \index{absorption imaging} \index{fluorescence imaging}

Besides projection-noise-limited resolution, another desirable feature of the clock's readout is that it be non-destructive.
Any trapped-atom clock will suffer from dead time while fresh atoms are cooled and loaded into the trap, an operation that normally takes several seconds.
During this dead time the noise of the local microwave oscillator is left uncorrected (Dick effect).
If a given sample of cold trapped atoms can be reused for multiple successive frequency measurements, then the clock can spend less of its time loading fresh atoms and more of it measuring frequency, suppressing local oscillator noise and improving stability. \index{Dick effect}
Furthermore, by reducing the number of new atoms loaded for each measurement, such reuse can also reduce the effects of trap loading noise, which is typically well above the atom-number shot noise.

\begin{figure}[htb]
\includegraphics[width=2.5in]{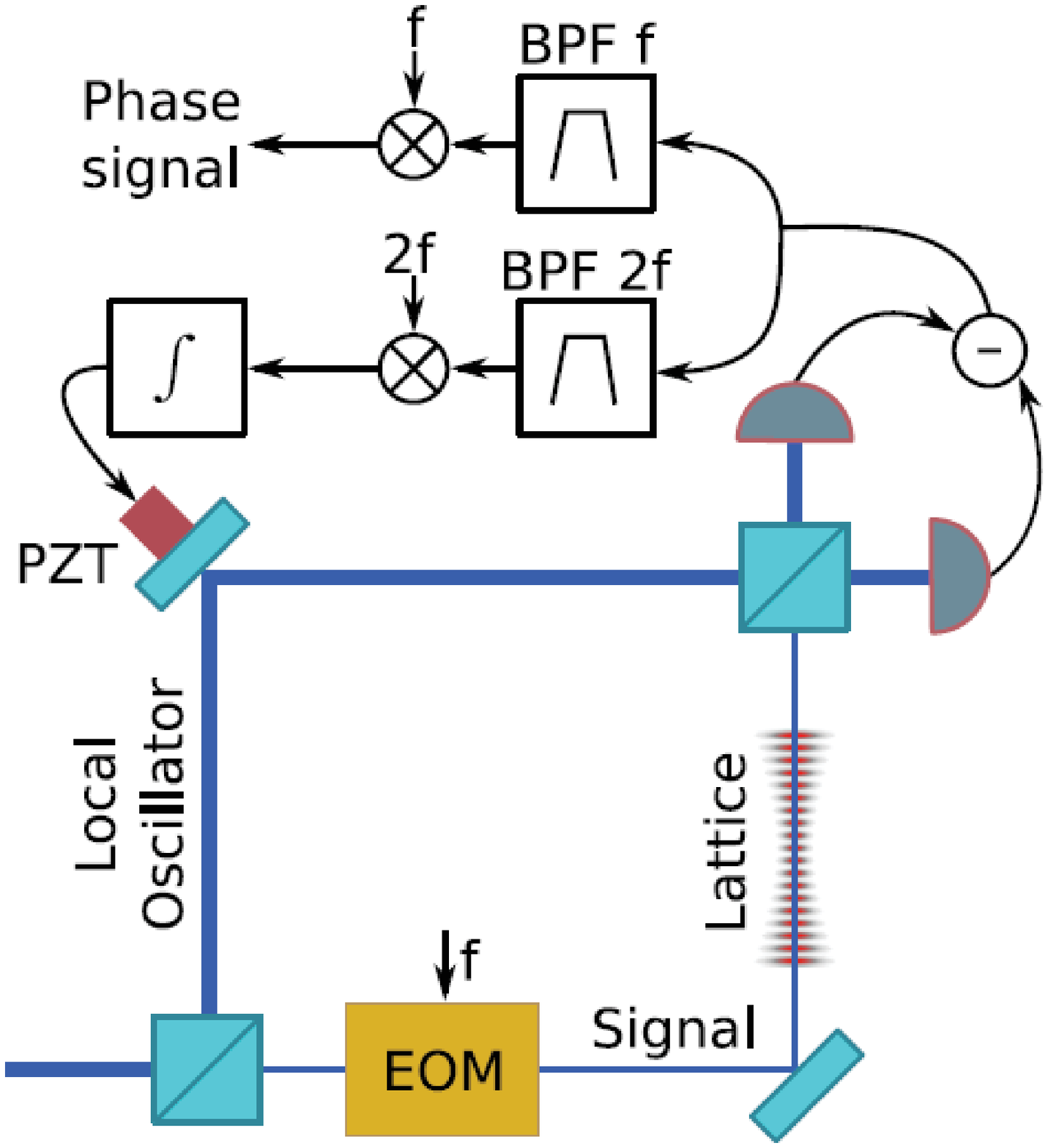}
\Caption{Setup for non-destructive detection of atomic-state populations in an $^{87}$Sr optical
lattice clock (from Ref. \cite{Vuletic:Lodewyck09}). The phase shift of the RF component at the modulation frequency $f$ is proportional to the number of atoms in the optical lattice.
The harmonic at frequency $2f$ is used to lock the phase of the interferometer. This setup
allows one to measure atomic-state populations with high signal-to-noise ratio without
losing the atoms from the lattice. This enables the repeated use of the same atoms for
several clock cycles, thereby improving the clock duty cycle and stability.} 
\label{Vuletic:fig:NondestructiveReadout}
\end{figure}

For this reason, Lodewyck \emph{et. al.} have built a non-destructive dispersive readout system into their optical clock apparatus \cite{Vuletic:Lodewyck09}.
They trap around $10^4$ \Sr\ atoms in the ground state of an optical lattice, obtaining a cloud with a radius of \unit[10]{\micro m}.
The trap is placed in one arm of a Mach-Zehnder interferometer with a probe beam waist of \unit[37]{\micro m}.
The weak ($\sim\unit{nW}$) probe beam undergoes a dispersive phase shift due to the presence of the atoms, which is detected by beating it against the strong ($\sim\unit{mW}$) optical local oscillator at the final beam-splitter of the interferometer.
The two output ports of this beam-splitter are detected on fast photodiodes with an overall quantum efficiency of 43\% and the signals subtracted, to obtain a signal that is insensitive to laser power drift to first order.
Rather than using the DC signal, which measures the phase between probe and local oscillator and would require careful stabilisation of the relative path length of the two interferometer arms, the detector makes use of a sideband modulation scheme reminiscent of the Pound-Drever-Hall frequency locking technique for optical resonators.
An electro-optic crystal in the probe arm of the interferometer modulates the probe beam phase at an RF frequency $f=\unit[90]{MHz}$, thus generating a pair of sidebands to either side of the atomic resonance, detuned by approximately three atomic linewidths.
Since the sidebands have opposite detuning, they are subjected to opposite dispersive phase shifts from the atoms, and this \emph{differential} phase shift is detectable as a phase shift of the RF component at frequency $f$ of the interferometer output.
Fluctuations in the path length difference of the interferometer impart a \emph{common} phase shift to the two sidebands and do not contribute to this signal to first order.
Path length fluctuations only affect the amplitude of the atomic signal, and a servo loop based on the $2f$ component of the output is used to detect such fluctuations and lock the interferometer at the position of maximum signal.
The net result is a photon-shot-noise limited determination of the dispersive atomic phase shift, yielding a measurement of the atom number in one of the two clock states at the projection noise level for $10^4$ atoms.
Up to 95\% of the atoms in the sample remain in the trap after such a readout and can be reused for the next cycle after a few ten \unit{ms} of recooling.
The authors foresee realistic clock operation with only $\unit[100]{ms}$ of dead time between Ramsey interrogations, improving the duty factor of the clock from the typical value of 10\% to over 80\%. \index{non-destructive clock readout} \index{duty factor}

\begin{figure}[htb]
\includegraphics[width=3.5in]{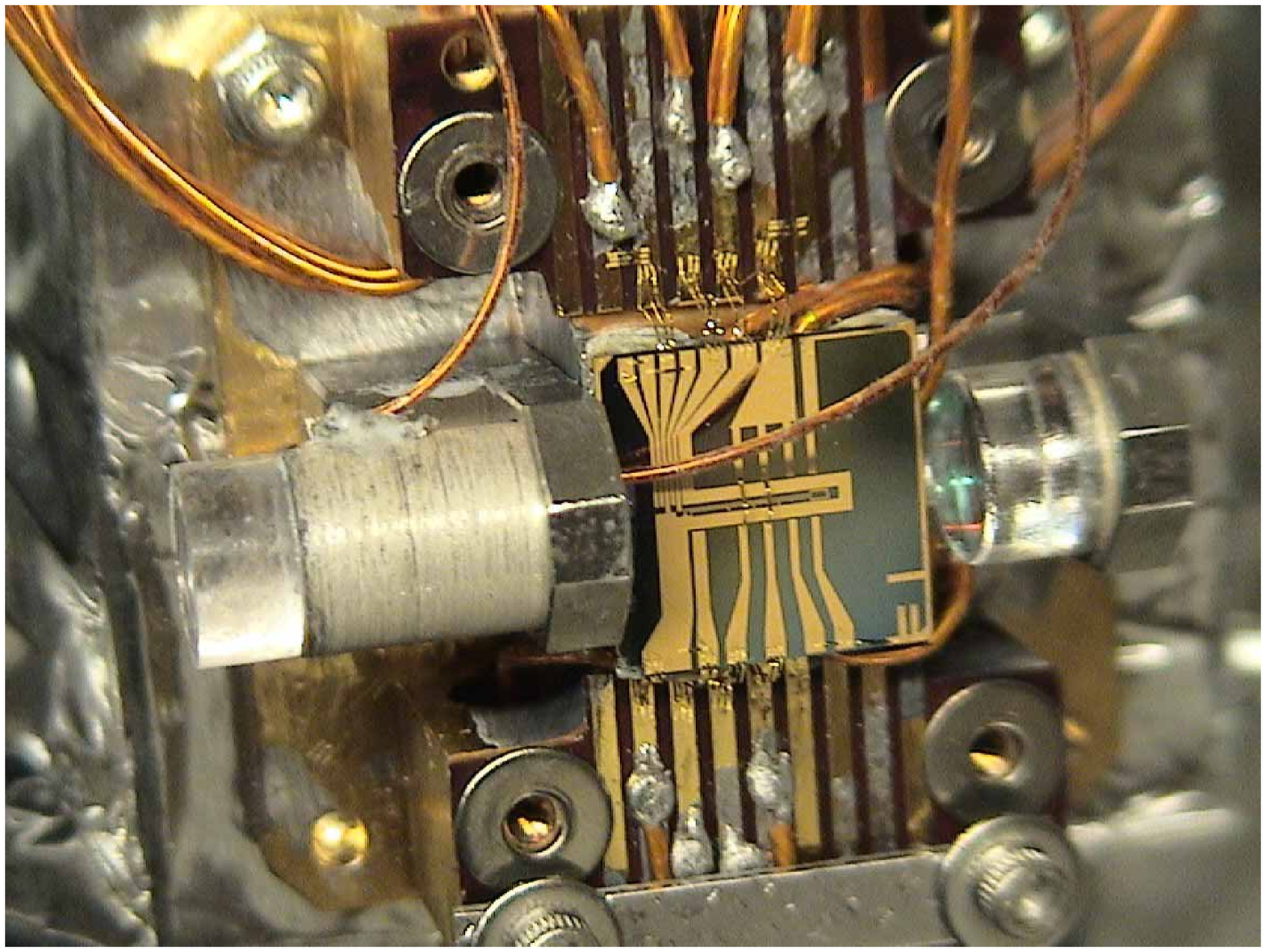}
\Caption{Optical resonator mounted on microchip \cite{Vuletic:Teper06,Vuletic:Schleier-Smith10,Vuletic:Leroux10} for quantum non-demolition measurements of the clock states and spin squeezing. The resonator mode is aligned at a height of 200$\mu$m above the chip.} 
\label{Vuletic:fig:Microchip}
\end{figure}

Another approach that has been successful in chip-based experiments employs dispersive state measurement \cite{Vuletic:Kuzmich98,Vuletic:Kuzmich00,Vuletic:Takeuchi07,Vuletic:Takano09,Vuletic:Takano10,Vuletic:Teper08,Vuletic:Koschorreck10} via an optical resonator, which has been used to realize both projection-noise-limited readout and spin squeezing (see section~\ref{Vuletic:sec:Squeezing}) for an atomic clock operated on a microchip \cite{Vuletic:Schleier-Smith10,Vuletic:Leroux10,Vuletic:Leroux10clock}.
In the demonstration of reference \cite{Vuletic:Schleier-Smith10}, an ensemble of up to $10^5$ \Rb\ atoms was prepared in an elongated trap overlapped with an optical resonator mode \unit[200]{\micro m} from the surface of a microchip, and transferred into a far-detuned optical trap formed inside the resonator (Fig. \ref{Vuletic:fig:Microchip}).
The deep standing-wave dipole trap ensured that the sample with radial extension of $\unit[8.1(8)]{\micro m}$, was located well within the resonator mode with a waist of $\unit[56.9(4)]{\micro m}$.
The cavity length was adjusted such that one optical resonance lay between the optical resonance frequencies for the two hyperfine clock states $\ket{F=1, m=0}$ and $\ket{F=2, m=0}$.
The state-dependent index of refraction of the atomic sample then changed the cavity resonance frequency by an amount proportional to the population difference between the clock states.
Measuring this shift provided a resolution of $\sim 30$ atoms, substantially better than the projection-induced fluctuations of over one hundred atoms.
Much less than one photon per atom was scattered into free space during the measurement so that the atomic sample was little heated and could be reused.

\section{Spin squeezing}
\label{Vuletic:sec:Squeezing}

Even in the absence of technical readout noise, the standard quantum limit of equation (\ref{Vuletic:eq:Precision}) places a bound on the achievable signal-to-noise ratio of a clock operated with an ensemble of $N$ independent atoms.
This fundamental instability---due to quantum projection noise on the final state readout---can in principle be made arbitrarily small by increasing the total atom number $N$.
In practice, however, the allowable atom number is severely limited by the onset of collision shifts which, as we have seen, are the bane of compact trapped-atom clocks.
Further improvements to chip clock stability must therefore be made at fixed $N$ by overcoming the projection noise limit.
This can be achieved by treating the $N$ atoms not as independent particles but as a single (entangled) ensemble, with quantum correlations between atoms.
Such correlations can be used to generate a Ramsey fringe that oscillates faster with frequency than a single-atom Ramsey fringe, increasing the size of the signal for constant projection noise.
This approach has been demonstrated in trapped-ion clocks \cite{Vuletic:Meyer01}, but it is not clear how to engineer the required entangled states in large ensembles such as those envisaged for chip-based neutral-atom clocks.
Alternatively, quantum correlations can be used to reduce the projection noise of the ensemble below the limit for independent particles, an approach known as spin squeezing.
Spin squeezing of superpositions of clock states is an active research area, with several recently demonstrated implementations \cite{Vuletic:Esteve08,Vuletic:Appel09,Vuletic:Schleier-Smith10,Vuletic:Leroux10,Vuletic:Gross10,Vuletic:Riedel10}, and more expected in the near future. \index{spin squeezing}

So far, three spin squeezing methods have been demonstrated in experiments.  
The first employs a non-destructive measurement of the ensemble to prepare it in a squezed initial state \cite{Vuletic:Kuzmich98,Vuletic:Bouchoule02,Vuletic:Appel09,Vuletic:Schleier-Smith10,Vuletic:Louchet-Chauvet10}.
The essential idea is to measure the quantum projection noise separately, before the clock is operated, so that it may be removed from the final Ramsey signal.
As long as less than one photon per atom is scattered into free space, measurement of the ensemble state with a single spatial mode of probe light does not reveal the states of the individual atoms, but merely how many are in each of the clock levels.
Therefore a measurement that resolves the atom number in each clock state to better than the projection noise limit can entangle the atoms, where the entangled state of the ensemble is conditioned on the outcome of the measurement.
The outcome of a later readout can then be predicted more precisely than would be possible for independent particles.
The cavity-enhanced state readout described in section \ref{Vuletic:sec:Readout} has been used to implement such measurement-based squeezing.
By comparing this reduction of readout noise to the reduction of coherence (and hence clock signal) inevitably induced by scattering and dephasing during the first measurement, an enhancement of the signal to noise ratio by $\unit[3.0(8)]{dB}$ was demonstrated.
The group of Eugene Polzik obtained similar results without the aid of a cavity by using a precision-stabilized Mach-Zehnder interferometer to detect the atomic index of refraction \cite{Vuletic:Appel09}.
They have used this technique as the basis for a prototype squeezed clock \cite{Vuletic:Louchet-Chauvet10}.

A second squeezing method relies on repulsive interactions between atoms to deterministically entangle them and produce states with reduced uncertainty on the population difference between the two clock states \cite{Vuletic:Gross10,Vuletic:Riedel10}.
Unlike measurement-based squeezing, this method unconditionally prepares the same squeezed state on every experimental cycle, which has the practical benefit that the squeezing is independent of the performance of the detection apparatus: one can know the initial squeezed state without having to observe it.
However, in the context of precision timekeeping, it has the salient drawback of relying upon the very same collisional energy shifts which are so detrimental to the clock's performance.
Thus, while this technique has a bright future in the study of many-body entanglement and may be useful for atom interferometry, it is unlikely to be of much use in improving the performance of chip clocks.

A third method has been developed that combines the benefits of the previous two techniques: deterministic squeezing by cavity feedback \cite{Vuletic:Takeuchi03,Vuletic:Schleier-Smith10a}.
In this method, the atoms are placed inside an optical resonator, as in the resonator-enhanced readout of section \ref{Vuletic:sec:Readout}.
However, instead of being projected into a squeezed state by measurement, the atomic ensemble evolves deterministically into a squeezed state due to the collective interaction of the atoms with the the light field of the resonator, as follows.
The atomic quantum noise tunes the resonator frequency just as a clock signal would, which for an incident light field tuned to the slope of the resonator line results in an intracavity intensity that depends on the atomic population difference.
The light-shift-induced phase evolution in each individual atom thus depends on the population difference of all atoms in the ensemble, leading to quantum correlations between atoms.
This approach generates spin dynamics similar to those of the one-axis twisting Hamiltonian in the original spin squeezing proposal of Kitagawa and Ueda \cite{Vuletic:Kitagawa93}, or to the repulsive interaction employed in the collisional squeezing experiments \cite{Vuletic:Gross10,Vuletic:Riedel10}.
However, since the incident light intensity can be switched to zero at will, the light-mediated interaction can be turned on only for the preparation of the initial squeezed state and then switched off to avoid perturbing the clock during the Ramsey precession time.
The cavity feedback method has produced the largest spin squeezing at the time of writing, a 5.6(6) dB improvement in signal-to-noise ratio \cite{Vuletic:Leroux10}.
It has also been used to operate a proof-of-principle clock which, for integration times up to \unit[50]{s}, achieved a stability 2.8(3) times better than the standard quantum limit of equation (\ref{Vuletic:eq:Precision}) \cite{Vuletic:Leroux10clock}.
The fractional stability of this clock was poor in concrete terms ($\sigma(\tau)=1.1\times10^{-9}\unit{s^{1/2}}/\sqrt{\tau}$), primarily because a very short Ramsey precession time of \unit[200]{\micro s} was used for the demonstration, 
but together with the experiment of reference \cite{Vuletic:Louchet-Chauvet10} it shows that even the standard quantum limit need not be an insuperable obstacle to the improvement of stability in chip clocks.

\setlength{\bibindent}{4mm} 

\bibliographystyle{atchip}
\bibliography{AtomChipsVuletic} 

\end{document}